\documentclass[11pt,oneside,a4paper]{article}
\pdfoutput=1   
\usepackage{amsmath,amsfonts,amssymb,amsbsy}
\usepackage{mathrsfs,latexsym}
\usepackage{graphicx}
\usepackage{color}
\usepackage{booktabs}
\usepackage{alpha}
\hypersetup{%
   pdftitle = {The b-quark mass from non-perturbative Nf=2 HQET at O(1/m)},
   pdfauthor = {F.Bernardoni, B.Blossier, J.Bulava, M.DellaMorte, P.Fritzsch, N.Garron, A.Gerardin, J.Heitger, G.vonHippel, H.Simma, R.Sommer},
   pdfkeywords = {Lattice QCD, Heavy Quark Effective Theory, b-quark mass}
}%
\usepackage{macros}
\usebiblio{hqet_nf2_mb}
\begin{document}

\preprintno{%
DESY 13-224\\
HU-EP-13/66\\
IFIC/13-82\\
LPT-Orsay/13-85\\
MITP/13-066\\
MS-TP-13-30\\
SFB/CPP-13-96\\
TCD-MATH-13-14\\
}

\title{%
The b-quark mass from non-perturbative $\Nf=2$ \\
Heavy Quark Effective Theory at $\Or(\minvh)$}

\collaboration{\includegraphics[width=2.8cm]{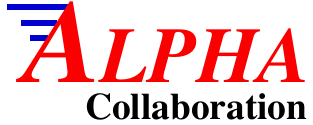}}

\author[desy]{Fabio~Bernardoni}
\author[fra]{Beno\^it~Blossier}
\author[trin]{John~Bulava}
\author[esp]{Michele~Della~Morte}
\author[hu]{Patrick~Fritzsch}
\author[trin]{Nicolas~Garron}
\author[fra]{Antoine~G\'erardin}
\author[wwu]{Jochen~Heitger}
\author[prism]{Georg~von~Hippel}
\author[desy]{Hubert~Simma}
\author[desy]{Rainer~Sommer}

\address[desy]{NIC @ DESY, Platanenallee~6, 15738~Zeuthen, Germany}
\address[fra]{Laboratoire~de~Physique~Th\'eorique, Universit\'e~Paris~XI,  91405~Orsay~Cedex, France}
\address[trin]{School~of~Mathematics, Trinity~College, Dublin~2, Ireland}
\address[esp]{IFIC and CSIC, Calle Catedr\'atico Jos\'e Beltran 2, 46980~Paterna, Valencia, Spain}
\address[hu]{Institut~f\"ur~Physik, Humboldt-Universit\"at~zu~Berlin, Newtonstr.~15, 12489~Berlin, Germany}
\address[wwu]{Institut~f\"ur~Theoretische~Physik, Universit\"at~M\"unster, Wilhelm-Klemm-Str.~9, 48149~M\"unster, Germany}
\address[prism]{PRISMA~Cluster~of~Excellence, Institut~f{\"u}r~Kernphysik, University~of~Mainz, Becherweg~45, 55099~Mainz, Germany}

\begin{abstract}
We report our final estimate of the b-quark mass from $\Nf=2$ lattice QCD
simulations using Heavy Quark Effective Theory non-perturbatively matched to
QCD at ${\Or}(\minvh)$.  Treating systematic and statistical errors in a
conservative manner, we obtain $\mbar_{\rm b}^{\msbar}(2\,\GeV)=4.88(15)\,\GeV$
after an extrapolation to the physical point.
\end{abstract}


\maketitle

\def\paperorletter{letter}

\section{Introduction}

The masses of the quarks are among the fundamental parameters of the Standard
Model (SM), and as such hold considerable interest.  Heavy quark masses, in
particular, enter as parameters in various perturbative predictions of
interesting decay rates, e.g.  $B \to X_{\rm s} \gamma$ or inclusive $B \to u$
or $B \to c$ rates. Such decays yield useful constraints for the CKM matrix
and, in principle, options to obtain hints for physics beyond the Standard
Model.  It is therefore desirable to minimize the uncertainty in $\mbeauty$
entering these predictions. 

The b-quark mass also enters the prediction for the cross section of the 
$H\to b\overline{b}$ decay, which is the  mode with the largest branching ratio 
for an SM-like Higgs with a mass of 126 GeV. In the future, tests of this 
coupling will help providing further characterizations of the new boson.

The most accurate determinations of the b-quark mass reported in the PDG review
\cite{Beringer:1900zz,Narison:2012xy,Bodenstein:2011fv,Hoang:2012us,Narison:2011rn,Laschka:2011zr,Aubert:2009qda,Chetyrkin:2009fv,Abdallah:2008ac,Schwanda:2008kw,Abdallah:2005cv,Boughezal:2006px,Buchmuller:2005zv,Pineda:2006gx,Bauer:2004ve,Hoang:2004xm,Bordes:2002ng,Corcella:2002uu,Eidemuller:2002wk,Erler:2002bu,Mahmood:2002tt,Brambilla:2001qk,Penin:2002zv,McNeile:2010ji,Dimopoulos:2011gx}
come from comparisons of experimental results for the $e^{+}e^{-} \to
b\overline{b}$ cross section to theoretical predictions from perturbation
theory and sum-rules.

Like each of these approaches, the first-principles determination of 
$m_{\rm b}$ from lattice field theory has its own difficulties.  Relativistic 
b-quarks cannot yet be reliably simulated on the lattice as their Compton 
wavelength is much shorter than any lattice spacing which can be currently 
reached in large-volume simulations.  To circumvent this limitation, two 
approaches have been used, {\it viz.} extrapolating simulation results 
obtained in the vicinity of the charm quark mass to the b-quark region
\cite{McNeile:2010ji,Dimopoulos:2011gx,deDivitiis:2003iy,Guazzini:2007ja,Carrasco:2013zta},
and the use of effective field theories, such as NRQCD
\cite{Gray:2005ur,Hart:2010jn}.
The approach of the ALPHA collaboration is based on Heavy Quark Effective
Theory (HQET)
\cite{stat:eichhill1,stat:symm1,stat:symm3,Eichten:1990vp},
which provides a description of heavy quarks in the context of heavy-light
mesons that can be employed in lattice QCD simulations if the
parameters of HQET are determined by matching HQET to QCD non-perturbatively
\cite{Heitger:2003nj,DellaMorte:2006cb}.
The matching at order $\Or(\minvh)$ has been performed in both the
quenched ($\Nf=0$) and the $\Nf=2$ theories by our collaboration
\cite{Blossier:2010jk,Blossier:2012qu}.
The lattice approach in general offers the unique opportunity to study the
$N_{\rm f}$-dependence of the b-quark mass.  We will further discuss that in
the conclusions.

In this \paperorletter, we present our results for the mass of the b-quark from
simulations of non-perturbative $\Nf=2$ HQET.  In Section \ref{s:methods}, we
briefly review the methods employed before presenting the results in
Section \ref{s:results}.  Section \ref{s:conclusions} contains our conclusions.

\section{Methodological Background}
\label{s:methods}

HQET on the lattice constitutes a theoretically sound approach to heavy quark
physics by expanding QCD correlation functions into power series in $\minvh$
around the static limit $m_{\rm h}\to\infty $, which is non-perturbatively
renormalizable, so that the continuum limit can always be taken.

Following the strategy described in
\cite{lat02:rainer, Heitger:2003nj}
and previously applied to calculate $\mb$ in the quenched approximation
\cite{DellaMorte:2006cb},
we write the HQET action at $\Or(\minvh)$ as
\begin{eqnarray}\label{e:hqetaction}
   S_{\rm HQET} &=& a^4\sum_x \left\{ \Lstat(x) - \omega_{\rm kin} \Okin(x) 
                                    - \omega_{\rm spin} \Ospin(x) \right\} \,, 
   \\           \label{e:lstat}
   \Lstat(x)    &=& \heavyb(x) \left(D_0 + \mhbare \right)\psi_{\rm h}(x)\,, 
   \\           \label{e:ofirst}
   \Okin(x)     &=& \heavyb(x) \vecD^2\psi_{\rm h}(x)\,,\\
   \Ospin(x)    &=& \heavyb(x) \vecsigma \cdot \vecB \psi_{\rm h}(x)\,,
\end{eqnarray}
where the heavy quark spinor field $\psi_{\rm h}$ obeys
$\frac{1+\gamma_0}{2}\psi_{\rm h}=\psi_{\rm h}$, $\mhbare$ is the bare heavy
quark mass absorbing the power-divergences of the self-energy in the static
approximation, and the parameters $\omega_{\rm kin}$ and $\omega_{\rm spin}$
are formally of order $\minvh$ and have been previously determined in
\cite{Blossier:2012qu}. 
The $\Or(\minvh)$ terms in \eqref{e:hqetaction} are treated as operator
insertions in static correlation functions:
\begin{equation}
\langle O\rangle = \langle O\rangle_{\rm stat} +
\omega_{\rm kin}\,a^4\sum_x \langle O\,\Okin(x)\rangle_{\rm stat} +
\omega_{\rm spin}\,a^4\sum_x \langle O\,\Ospin(x)\rangle_{\rm stat}
\end{equation}
for the expectation value of some multilocal fields $O$, where $\langle
O\rangle_{\rm stat}$ is the expectation value of $O$ determined in the static
theory.  A significant improvement in the signal-to-noise ratio of heavy-light
correlation functions can be achieved by defining the covariant backward time
derivative $D_0 f(x) = (f(x) - U^\dagger(x-a\hat0,0)f(x-a\hat0))/a$ in terms of
a suitably smeared link instead of the bare link $U(x,0)$.  Each smearing
prescription constitutes a separate lattice action; here we have employed both
the HYP1 and HYP2 actions
\cite{HYP,HYP:pot,DellaMorte:2005yc}.

To reliably extract hadronic quantities, we have to pay particular attention to
unwanted contributions from excited states.  The variational method, which has
become a standard tool for analyzing hadronic spectra in lattice QCD, starts
from correlator matrices
\begin{eqnarray}  \nonumber
  C^{\rm{stat}}_{ij}(t) &=& \sum_{x,\bf{y}} \left< O_i(x_0+t,{\bf y})\,O^*_j(x)\right>_\stat\,,
  \\[-0.7em]            \label{e:cmatdefs}
  \\[-0.2em]            \nonumber
  C^{\rm{kin/spin}}_{ij}(t)& = & \sum_{x,\vecy,z}\;
  \left< O_i(x_0+t,\vecy)\,O^*_j(x)\,  {\cal O}_{\rm kin/spin}(z)\right>_\stat \,,
\end{eqnarray}
for a suitably chosen basis of interpolating fields $O_i\,,\;i=1,\ldots,N$. The
main ingredient is to solve the generalized eigenvalue problem (GEVP) in the
static limit
\begin{eqnarray} \label{e:gevp}
   C^{\rm stat}(t)\, v^{\rm stat}_n(t,t_0) = \lambda^{\rm stat}_n(t,t_0)\, C^{\rm stat}(t_0)\,v^{\rm stat}_n(t,t_0) \,,
   \quad n=1,\ldots,N\,,\quad t>t_0\;.
\end{eqnarray}
Indeed, by exploiting the orthogonality property of the eigenvectors 
\begin{equation}  \nonumber
   ( v_m^\stat(t,t_0), C^\stat(t_0)  v_n^\stat(t,t_0) ) \propto \delta_{nm}
\end{equation} 
one can show that the $\Or(\minvh)$ corrections to the energy levels depend
only on the static generalized eigenvalues $\lambda_n^\stat(t,t_0)$, the
eigenvectors $v_n^\stat(t,t_0)$, and the $\Or(\minvh)$ correlators 
$C^{\rm kin/spin}(t)$ \cite{Blossier:2009kd}, in analogy with perturbation 
theory in quantum mechanics. 
At large times $t$ and $t_0$ satisfying $t_0\ge t/2$, the asymptotic behaviour
is then known to be
\cite{Blossier:2009kd}
\begin{eqnarray}
    E_n^{\rm eff,\stat}(t,t_0) &=&
    E_n^\stat \,+\, \beta_n^\stat\,\rme^{-\Delta E_{N+1,n}^\stat\, t}+\ldots\,,
   \label{e:lamstatfit}
  \\[0.5em]
    E_n^{\rm eff,\first}(t,t_0) &=&
    E_n^\first \,+\, [\,\beta_n^\first
               \,-\, \beta_n^\stat\,t\,\Delta E_{N+1,n}^\first\,]
                     \rme^{-\Delta E_{N+1,n}^\stat\, t}+\ldots\, ,\qquad
   \label{e:lamfirstfit}
\end{eqnarray}
with $\Delta E_{m,n}=E_m-E_n$. The time intervals over which we fit the energy
plateaux are chosen so as to minimize the systematic error from the excited
states while keeping the statistical error under control.

Finally, the mass of the B-meson to $\Or(\minvh)$ is given by ($E^{\rm x}
\equiv E_1^{{\rm x}}$)
\begin{eqnarray}\label{eq:mB_nlo}
  m_{\rm B} = \mhbare + E^\stat + \omegakin E^{\rm kin} +\omegaspin E^{\rm spin} \,
\end{eqnarray}
and it remains to perform the chiral and continuum extrapolation and to solve
the equation $m_{\rm B}(m_{\rm h}=m_{\rm b})=m^{\rm exp}_{\rm B}$ by an
interpolation.

\section{Simulation details and results}
\label{s:results}

\subsection{Ensembles used}

Our measurements are carried out on a subset of the CLS (Coordinated Lattice
Simulations) ensembles, which have been generated using either the DD-HMC
\cite{algo:L1a,algo:L2,algo:L3,soft:DDHMC}
or the MP-HMC
\cite{Marinkovic:2010eg}
algorithm, using the Wilson plaquette action
\cite{Wilson}
and $\Nf=2$ flavours of non-perturbatively $\Or(a)$ improved Wilson quarks
\cite{impr:SW,impr:pap3}.
An overview of the simulation parameters of the ensembles used is given in
Table~\ref{tab:parameters}. In order to suppress finite-size effects, we
consider only ensembles satisfying $m_{\pi} L > 4.0$.  The light valence quarks
are equal to the sea quarks, and the (quenched) b-quark is treated by HQET.

\begin{table}[t]
  \centering\small
  \begin{tabular}{@{\extracolsep{0.1cm}}ccccccccc}
  \toprule
  $\beta$ & $a$[fm] & $L/a$ & $m_\pi$[MeV] & $m_\pi L$ & \#cfgs &  $\dfrac{\#\rm cfgs}{\tau_{\rm exp}}$  & id  & $\{R_1,R_2,R_3\}$ \\\midrule     
     $5.2$ &$0.075$ & $32$  & $380$        & $4.7$     &  1012  &  122  & A4 & $\{15,  60, 155\}$ \\
           &        & $32$  & $330$        & $4.0$     &  1001  &  164  & A5   \\
           &        & $48$  & $280$        & $5.2$     &   636  &  52   & B6   \\ \midrule
     $5.3$ &$0.065$ & $32$  & $440$        & $4.7$     &  1000  &  120  & E5 & $\{22,  90, 225\}$ \\
           &        & $48$  & $310$        & $5.0$     &   500  &  30   & F6   \\
           &        & $48$  & $270$        & $4.3$     &   602  &  36   & F7   \\
           &        & $64$  & $190$        & $4.1$     &   410  &  17   & G8   \\\midrule
     $5.5$ &$0.048$ & $48$  & $440$        & $5.2$     &   477  &  4.2  & N5 & $\{33, 135, 338\}$ \\
           &        & $48$  & $340$        & $4.0$     &   950  &  38   & N6   \\
           &        & $64$  & $270$        & $4.2$     &   980  &  20   & O7   \\
  \bottomrule
  \end{tabular}
  \caption{Details of the CLS ensembles used: bare coupling $\beta=6/g_0^2$, 
           lattice spacing $a$, spatial extent $L$ in lattice units ($T=2L$),
           pion mass $m_\pi$, $m_\pi L$, number of configurations employed,
           and number of configurations employed normalized in units of
           the exponential autocorrelation time $\tau_{\rm exp}$ as 
           estimated in \cite{Schaefer:2010hu}. Additionally, we specify
           the CLS label id and the Gaussian smearing parameters $R_k$ used 
           to build different interpolating fields as described in the text.
           } 
  \label{tab:parameters}
\end{table}

In order to control the statistical error in a reliable fashion, we make use of
the method of
\cite{Wolff:2003sm}
as improved by
\cite{Schaefer:2010hu}
to estimate the effect of long-term autocorrelations due to the coupling of our
observables to the slow modes of the Markov chain, decaying as 
$\sim \exp(-\tau/\tau_\mathrm{exp})$ in Monte Carlo simulation time $\tau$. The
propagation of these effects through to the continuum-extrapolated result at
the physical pion mass is carried out iterating the formulae of
\cite{Schaefer:2010hu}.

\begin{figure}[t]
  \centering
  \includegraphics[width=\textwidth]{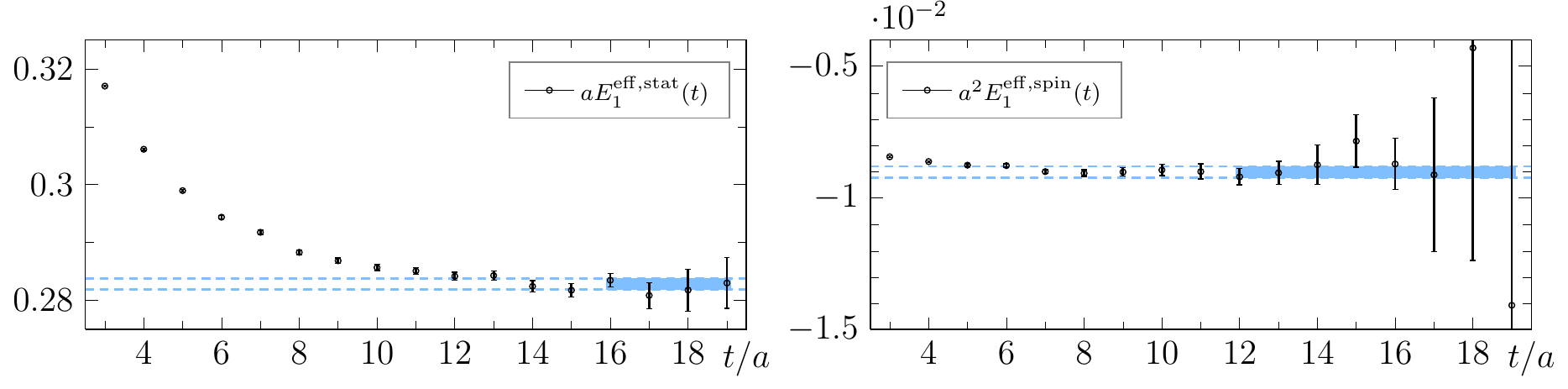}
  \caption{Illustration of typical plateaux for the ground state 
           static energy (left panel) and the $\Or(\minv )$ 
           chromomagnetic energy (right); the CLS ensemble shown 
           here is N6 ($a=0.048$\,fm, $m_{\pi}=340$\,MeV).
          }
  \label{fig:plateaux}
\end{figure}
%
%

\subsection{Lattice spacings}

The lattice spacings $a$, pion masses $\mpi$ and pion decay constants $\fpi$ on
the CLS ensembles used here are taken from an update 
\cite{proceedingsLottini}
of the analysis in
\cite{Fritzsch:2012wq}
with increased statistics and including additional ensembles.  They read
\begin{align}
  a &= 0.04831(38) \,\fm  & \text{at}~ \beta=5.5\,, \nonumber \\
  a &= 0.06531(60) \,\fm  & \text{at}~ \beta=5.3\,, \\
  a &= 0.07513(79) \,\fm  & \text{at}~ \beta=5.2\,, \nonumber
\end{align}
and result from setting the scale via the kaon decay constant, $\fK=155$\,MeV.
With the updated values of $a\fK$ we follow the lines of~\cite{Fritzsch:2012wq}
and re-evaluate
\begin{align}
   L_1 \fK &= \lim_{a\to 0}\, [L_1/a][a\fK] = 0.312(8)     \label{eq:L1fK} 
\end{align}
that is needed to convert the b-quark mass into physical units later on.  The
length scale $L_1$ originates from the non-perturbative finite-volume matching
step used to determine the HQET parameters~\cite{Blossier:2012qu}.

\subsection{Basis of B-meson interpolating fields}

Our basis of $N=3$ operators is given by
\begin{equation}
   O_k(x) = \psibar_{\rm h}(x)\gamma_0\gamma_5\psi_{\rm l}^{(k)}(x) \,,  
            \qquad k=1,\ldots,N\,,
\end{equation}
where $\psi_{\rm h}(x)$ is the static quark field, and different levels of
Gaussian smearing
\cite{Gusken:1989ad}
with a triply (spatially) APE smeared
\cite{smear:ape,Basak:2005gi}
covariant Laplacian $\Delta$ are applied to the relativistic quark field
\begin{equation}
   \psi_{\rm l}^{(k)}(x) = \left( 1+\kappa_{\rm G}\,a^2\,\Delta \right)^{R_k} 
                                  \psi_{\rm l}(x)\,.
\end{equation}
Our smearing parameters $\kappa_{\rm G}=0.1$ and $R_k$, collected in Table
\ref{tab:parameters}, are chosen so as to use approximately the same sequence
of physical radii $r_k=2a\sqrt{\kappa_{\rm G} R_k}$ at each value of the lattice
spacing.  In extracting our estimates for the energies 
$E_1^{\stat,\,{\rm kin},\, {\rm spin}}$ from the GEVP, the time intervals 
$[t_{\rm min}, t_{\rm max}]$ over which we fit the plateaux are chosen so as 
\begin{equation}
  r(t_{\rm min}) = \frac{|A(t_{\rm min})-A(t_{\rm min}-\delta)|}
                  {\sqrt{\sigma^2(t_{\rm min})+\sigma^2(t_{\rm min}-\delta)}} 
                   \le 3\,,
\end{equation}
where $A$ is the plateau average, $\sigma$ is the statistical error, 
$\delta = 2/(E^{\stat}_{N+1}-E^{\stat}_1)\sim 0.3\,\fm$, and $t_{\rm max}$ is 
fixed to $\sim 0.9\,\fm$.  This will assure that our selection criterion 
$\sigma_{\rm sys}\leq \sigma/3$ is satisfied \cite{Blossier:2010vz}, where 
$\sigma_{\rm sys}\propto\exp[-(E_{N+1}-E_1) t_{\rm min}]$.  An illustration 
of two typical plateaux of $E_1^{\stat}$ and $E_1^{\rm spin}$ is shown in
Fig.~\ref{fig:plateaux}.

\subsection{Determination of the b-quark mass}

The mass of the B-meson to static order is given by
\begin{eqnarray}  \label{eq:mB_stat}
  m_{\rm B}^\stat &=& \mhbare^\stat + E^\stat 
\end{eqnarray}
while the main formula at $\Or(\minv)$ was given in eq.~\eqref{eq:mB_nlo}.  The
HQET parameters $\mhbare^\stat$, $\mhbare$, $\omega_{\rm kin}$ and 
$\omega_{\rm spin}$ depend on the renormalization group invariant (RGI) heavy 
quark mass $M$ (defined below) and the lattice spacing $a$.  We parameterize 
this dependence by the dimensionless variable $z=M L_1$ and $a$, where $L_{1}$ 
is kept fixed. It is implicitly defined by the renormalized coupling in the 
Schr\"odinger functional (SF) scheme via $\bar{g}^{2}(L_{1}/2)=2.989$
\cite{Blossier:2012qu}.
Apart from $a$, the large-volume observables $E^{\rm x}$ depend on the light
quark mass which we parameterize through $\mpi$. Thus $m_{{\rm B},\delta}$,
computed with discretization HYP1 for $\delta=1$ and  HYP2 for $\delta=2$, are
functions of $z$, $m_\pi$ and $a$. Their values are listed in
Table~\ref{tabmB}.

\begin{table}[t]
   \centering\small
   \begin{tabular}{clccccccccccc} \toprule
       &     & \multicolumn{2}{c}{$z=11$}& \multicolumn{2}{c}{$z=13$}  & \multicolumn{2}{c}{$z=15$} \\ 
       \cmidrule(lr){3-4}\cmidrule(lr){5-6}\cmidrule(lr){7-8}
    id & $y$        & HYP1     & HYP2     & HYP1     & HYP2     & HYP1     & HYP2     \\ \midrule
    A4 & 0.0771(14) & 4434(62) & 4454(62) & 5024(70) & 5042(70) & 5597(78) & 5613(78) \\
    A5 & 0.0624(13) & 4419(62) & 4440(62) & 5010(70) & 5028(70) & 5583(78) & 5600(78) \\
    B6 & 0.0484(9)  & 4398(62) & 4420(62) & 4988(70) & 5008(70) & 5562(78) & 5579(78) \\[0.15em]
    E5 & 0.0926(15) & 4474(59) & 4492(59) & 5069(66) & 5084(66) & 5646(73) & 5661(73) \\
    F6 & 0.0562(9)  & 4436(59) & 4452(58) & 5031(66) & 5046(66) & 5609(73) & 5622(73) \\
    F7 & 0.0449(7)  & 4431(58) & 4444(58) & 5026(65) & 5037(65) & 5603(73) & 5613(73) \\
    G8 & 0.0260(5)  & 4415(59) & 4434(59) & 5010(66) & 5027(66) & 5589(73) & 5603(73) \\[0.15em]
    N5 & 0.0940(24) & 4586(57) & 4594(57) & 5193(64) & 5200(63) & 5783(71) & 5789(70) \\
    N6 & 0.0662(10) & 4563(57) & 4568(56) & 5169(63) & 5174(63) & 5759(70) & 5763(70) \\
    O7 & 0.0447(7)  & 4539(56) & 4555(56) & 5147(63) & 5161(63) & 5737(69) & 5750(70) \\
    \midrule
$B(z)$ &     &\multicolumn{2}{c}{ 4610(57) }& \multicolumn{2}{c}{ 5207(63) }& \multicolumn{2}{c}{ 5787(69) }\\
    \bottomrule
\end{tabular}

\
   \caption{\label{tabmB}
            Raw data of $\mBd(z,\mpi,a)$ in MeV for all ensembles (id), $z$ 
            and HYP actions considered in this work. In the last row we report
            $B(z)\equiv \mBd^{\rm sub}\left(z,m_\pi^{\rm exp},0\right)$ for 
            the $z$ that were used in the quadratic interpolation to fix $\zb$
            using eq.~\eqref{zbcond}.
        }
\end{table}
\begin{figure*}[t]
   \centering
   \includegraphics[height=0.27\textheight]{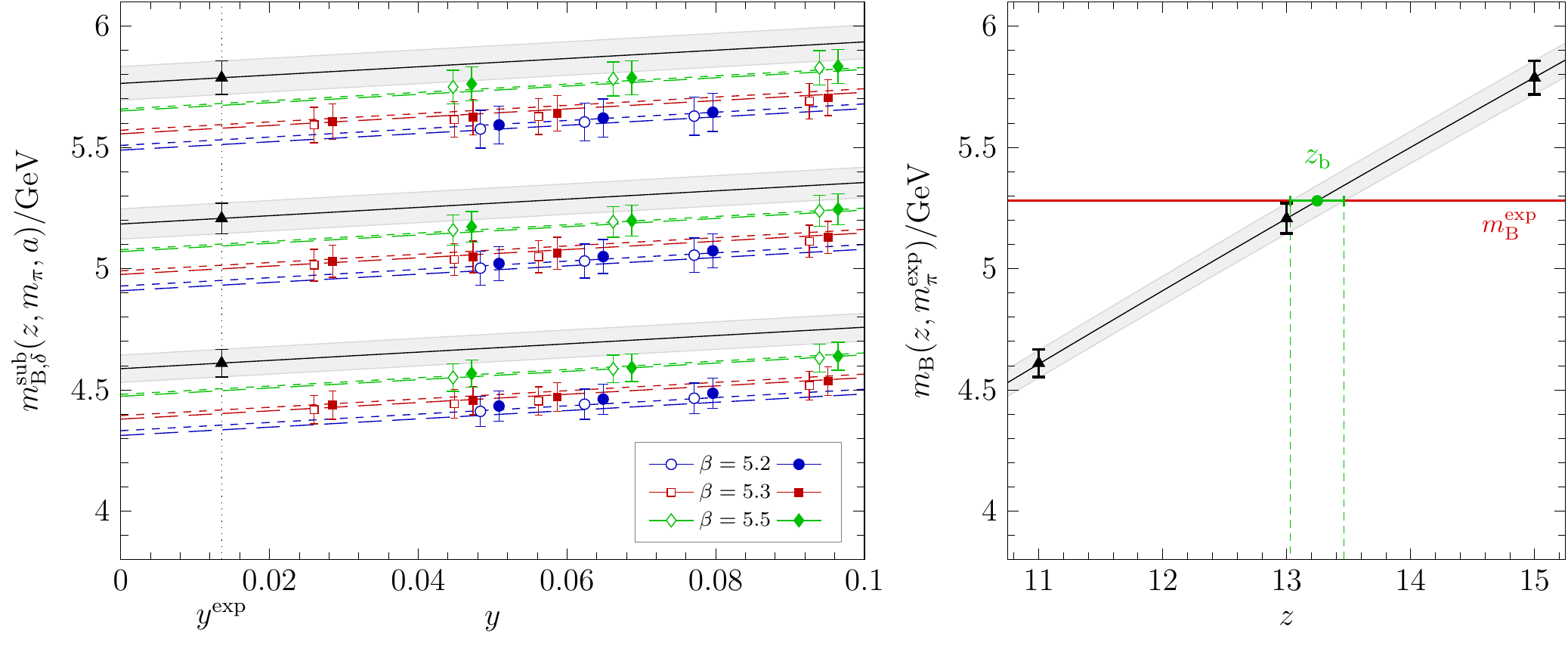}
   \caption{\label{figextrap}
            ({\it Left}) Chiral and continuum extrapolation of 
            $\mBd^{\rm sub}\left(z,y,a\right)$ for the $z$ used in the 
            determination of $\zb$. Open/filled symbols refer to HYP1/HYP2 data 
            points as do long/short dashed curves, respectively. 
            ({\it Right}) Interpolation to $\zb$ by imposing eq.~\eqref{zbcond}.
           }
\end{figure*}

Once $m_{{\rm B},\delta}$ have been computed for a set of $z$ spanning
a range of heavy quark masses containing the b-quark mass, we perform a
combined chiral and continuum extrapolation to obtain
$m_{{\rm B}}(z,\mpi^{\rm exp})\equiv m_{{\rm B},\delta}(z,\mpi^{\rm exp},0)$,
using $\mpi^{\rm exp}=134.98$\,\MeV~\cite{Beringer:1900zz}. Considering that 
the $\Or(a)$ improvement was performed non-perturbatively but neglecting 
$\Or(a/\mb)$ effects,%
\footnote{Accounting for an $a/\mb$ has little effect. Adding a term 
$F_\delta\cdot(a/\mb)$ to eq.~\eqref{chimB} does not change the unnormalized 
$\chi^2$. For instance, the fitting parameter $B(z)|_{z=13}$ changes to 
$5227(79)\,\MeV$ and eq.~\eqref{zb} would read $\zb=13.18(27)(13)$.
}
the NLO formula from HMChPT reads
\cite{Bernardoni:2009sx}
\begin{align}  \label{chimB}
   \mBd^{\rm sub}\left(z,y,a\right) 
        &= B(z) + C \left(y-y^{\rm exp} \right) + D_{\delta} a^2 \,, &
      y &\equiv \frac{m^2_\pi}{8\pi^2f_\pi^2} \;,
\end{align}
where in
\begin{align}  \label{defmsub}
   \mBd^{\rm sub}\left(z,y,a\right) 
        &\equiv \mBd\left(z,m_\pi,a\right)+ \frac{3\widehat{g}^2}{16\pi }
                \left( \frac{m^3_{\pi}}{f_{\pi}^2} - 
                \frac{(m_\pi^{\rm exp})^3}{(f_\pi^{\rm exp})^2} \right)
\end{align}
the leading non-analytic term of  HMChPT has been subtracted.  The $B^*B\pi$
coupling $\widehat{g} = 0.489(32)$ has been determined recently
\cite{Bulava:2010ej}
and the variable $y$ is identical to $\tilde{y}_1$ introduced in
\cite{Fritzsch:2012wq}.
We use the convention where the pion decay constant is 
$f_\pi^{\rm exp}=130.4\,\MeV$.  The extrapolation \eqref{chimB} is shown in
Fig.~\ref{figextrap} (left) for three values of $z$ in the vicinity of 
$\zb=\Mb L_1$ we are aiming at. Its result, $B(z)=\mBd(z,m_\pi^{\rm exp},0)$, 
is given in Table~\ref{tabmB}.  As shown in Fig.~\ref{figextrap} (right) the
corresponding dependence of $\mB$ on $z$ at the physical point is nearly
linear, indicating that the HQET expansion is precise for this observable.
Nevertheless, we perform a quadratic interpolation of 
$\mB\left(z,m_\pi^{\rm exp},0\right)$ and fix $\zb$ by imposing the 
experimental value for the B-meson mass,
\begin{equation}  \label{zbcond}
   \mB(z,m_{\pi}^{\rm exp},0)\big|_{z=\zb} \equiv \mB^{\rm exp} \,.
\end{equation}
We take $\mB^{\rm exp}=5.2795\,\GeV$~\cite{Beringer:1900zz} and obtain
\begin{equation}
   \zb = 13.25(22)(13)_z \,,     \label{zb}
\end{equation}
where the first error is statistical and in particular contains the error from
the combined chiral and continuum extrapolation, whereas the second error is
the uncertainty of $h(L_0)$ defined in \eqref{eq:hk}. It is due to the
non-perturbative quark mass renormalization in QCD
\cite{Blossier:2012qu}.
To give the RGI b-quark mass in physical units we combine~\eqref{zb}
and~\eqref{eq:L1fK} to solve the relation $\zb=L_1\Mb$ for $\Mb$.  According 
to $\Mb=\zb/[L_1\fK]\!\cdot\!\fK$ we finally obtain our main result%
\footnote{%
We follow the notation of Gasser and Leutwyler~\cite{Gasser:1982ap} for 
the definition of the RGI mass,
$  M = \lim_{\mu \to \infty} \left(2 b_0 \gbar^2(\mu)\right)^{-d_0/(2b_0)} \,\mbar(\mu) $, 
where $b_0=(11 - 2\Nf/3)(4\pi)^{-2} $ and $d_0=8 (4\pi)^{-2}$.
}
\begin{align}     \label{eq:Mb}
    \Mb &= 6.58(17) \,\GeV \;.
\end{align}
Since in the literature it is more common to compare masses in the  $\MSbar$
scheme, we convert our result~\eqref{eq:Mb} and give its value $\mbbMS$ at the
scale $\mu=\mbbMS$ as well as at $\mu=2\,\GeV$.  We use
\begin{align}  \label{eq:MSconv}
   \mbbMS(\mbbMS) = \Mb\cdot \rho(\Mb/\lMSbar) \,,
\end{align}
with a conversion function $\rho(r)$ that can be evaluated accurately
using the known 4-loop anomalous dimensions of quark masses and coupling
\cite{Chetyrkin:1999pq, Melnikov:2000qh}.
It is described in more detail in appendix~\ref{sec:a1}.  The ratio
$r_\mathrm{b}=\Mb/\lMSbar$ is computed from our value of $z_\mathrm{b}$ and the
ALPHA collaboration results for non-perturbative quark mass renormalization
\cite{DellaMorte:2005kg}.  We find $r_\mathrm{b}=21.1(13), \,
\rho(r_\mathrm{b})=0.640(6)$ and
\begin{align}   \label{eq:mbMS}
   \mbbMS(2\,\GeV) &=  4.88(15)\, \GeV\,, &
   \mbbMS(\mbbMS)  &=  4.21(11)\, \GeV\,.
\end{align}
We emphasize that this is the mass in the theory with two dynamical
quark-flavours, the b-quark is quenched, a completely well defined
approximation for a heavy quark. In particular also the function $\rho(r)$
refers to $\Nf=2$.

To have an idea of the magnitude of $\Or(\minv)$ corrections to the b-quark
mass one must repeat the above computation in the static limit. The reason
is that the $1/m$ contribution $\omegakin E^{\rm kin}+\omegaspin E^{\rm spin}$ 
is divergent in the continuum limit; only the combination with $\mhbare$ in 
eq.~\eqref{eq:mB_nlo} is finite. The HQET parameter $\mhbare^\stat$ was 
determined by matching the static theory with QCD as described in
\cite{Blossier:2012qu}.
By repeating the same steps as for the NLO case we obtain
\begin{align}      \label{eq:Mbstat}
  \zb^\stat &= 13.24(21)(13)_z  \,, &
  \Mb^\stat &= 6.57(17)\,  \GeV \,,
\end{align}
which after conversion to the $\MSbar$ scheme gives
\begin{equation}   \label{eq:mbstatMS}
  \left[ \mbbMS(\mbbMS) \right]^\stat = 4.21(11) \,\GeV\,.
\end{equation}
The result of the combined chiral and continuum extrapolation of $\mB$ in the
static limit, as well as the quadratic interpolation in $z$ to obtain
$\zb^\stat$ are very similar to those obtained at next-to-leading order.  The
small differences observed between the results in~\eqref{zb}--\eqref{eq:Mb}
and~\eqref{eq:Mbstat} show that for this observable the HQET expansion is very
precise, making us confident that $\Or(\minv^2)$ corrections are negligible
with present accuracy. Indeed, the smallness of the $\minv$ terms is known with
much higher accuracy than \eqref{zb} suggests, e.g.,
$\zb^\first\equiv\zb-\zb^\stat=-0.008(51)$.

We conclude this section by analyzing the error budget for $\zb$.  As can be
seen in Table~\ref{errbudget} approximately $62\%$ of the contribution to the
square of the error comes from the HQET parameters.  Another $\sim 21\%$ comes
from the relativistic $\za$ that affects the computation of $\zb$ through the
scale setting, while only the residual $\sim 17\%$ comes from the computation
of the HQET matrix elements. In this respect the largest contribution comes
from the ensembles at $\beta=5.5$, that are more affected by long-term
autocorrelations (critical slowing down).

\begin{table} 
 \small
 \centering
 \begin{tabular}{lccccccccc}\toprule
   source i                        & A3  & G8  & N5  & N6  & O7  & $\za$ & $\omega^{\rm HQET}$  \\\midrule
   $(\sigma_{\rm i}/\sigma)^2$[\%] & 1.2 & 0.9 & 2.6 & 5.9 & 5.6 & 20.6  & 61.6                 \\ 
   \bottomrule
 \end{tabular}
  \caption{\label{errbudget}
           Partial contributions $(\sigma_{\rm i}/\sigma)^2$ to the accumulated 
           error $\sigma$ of $\zb$. Only error sources contributing with a 
           relative squared uncertainty $(\sigma_{\rm i}/\sigma)^2>0.5\%$ are 
           listed. The ensemble A3 did not appear in table~\ref{tab:parameters} 
           since it enters through the scale setting 
           procedure~\cite{proceedingsLottini,Fritzsch:2012wq} only.
          }
\end{table}

\section{Discussion and conclusions} \label{s:conclusions}

Using non-perturbatively matched and renormalized HQET in $\Nf=2$ lattice QCD,
we have determined the mass of the b-quark with essentially controlled
systematic errors: in particular, the renormalization is carried out without
recourse to perturbation  theory and the continuum limit is taken.  An
irreducible systematic error which remains is a 
${\Delta \mb\,  /\, \mb}  \sim (\Lambda/\mb)^3$ relative error due to the 
truncation of the HQET expansion at order $\Lambda^2/\mb$. However, with a
typical scale of $\Lambda=500~\MeV$ one obtains a permille-sized truncation
error, which is completely negligible with today's accuracy.  The estimate is
supported by the fact that we do not see any difference between our static
result and the one including the $\Lambda^2/\mb$ terms.  Furthermore, according
to previous experience an effective scale of around $\Lambda=500~\MeV$ seems to
govern the expansion 
\cite{Blossier:2010vz,Blossier:2010mk,Blossier:2012qu}.

Our results,
\begin{align}
                 \Mb\big|_{\Nf=2} &= 6.57(17)  \,\GeV \,,  \\
     \mbbMS(2\,\GeV)\big|_{\Nf=2} &= 4.88(15)  \,\GeV \,,
\end{align}
are in agreement with the $\Nf=2$ results of \cite{Carrasco:2013zta} who cite a
similar error, but use a completely different approach.  We compare to the
quenched approximation and to the PDG values in \tab{masses}. There is little
dependence of $\mbbMS(\mu)$ on the number of flavours for $\Nf =0,2,5$ and for
typical values of $\mu$ between $\mbbMS$ itself and $2\,\GeV$. 

In particular at the lower scale of $2\,\GeV$, where the apparent convergence
of perturbation theory is still quite good, a flavour number dependence of the
mass of the b-quark is not detectable at all. In hindsight, this is rather
plausible as we match our effective theories (albeit with only $\Nf=0,2$
dynamical flavours) to the real world data at low energies.  Indeed, precisely
speaking the above statements refer to the theories renormalized by fixing the
B-meson mass to its physical value and setting the overall energy scale through
the kaon decay constant\cite{Fritzsch:2012wq} or roughly equivalent the pion
decay constant \cite{proceedingsLottini}.\footnote{For $\Nf=0$ we used the
scale $r_0\approx 0.5\,\fm$ instead of the decay constants, but in
\cite{Fritzsch:2012wq} this value of $r_0$ was obtained for the $\Nf=2$
theory.} In this way the low energy hadron sector of the theories is matched to
experiment, and it is natural to expect that the quark masses agree at a
relatively low scale. On the other hand we do not want to push the perturbation
theory needed for giving $\mbb$ in the $\MSbar$ scheme to scales below
$2\,\GeV$. We remark that also the strange quark mass at  $2\,\GeV$ is known to
be only weakly dependent on $\Nf$ \cite{FLAG11,FLAG13}.

In contrast, the RGI mass $\Mb$ differs significantly between $\Nf=5$ and
$\Nf=2$.  Given the observed weak flavour number dependence at scales of 2-5
GeV, the differences in $\Mb$ can be traced back to the $\Nf$ dependence of
both the RG functions and the $\Lambda$ parameters. These two effects happen to
reinforce each other between $\Nf=5$ and $\Nf=2$ while in the comparison
$\Nf=2$ and $\Nf=0$ they partially compensate.

All of this suggests to use the b-quark mass at scales around $\mu=2$~GeV when
one attempts to make predictions from theories with a smaller number of
flavours for the physical 5-flavour theory. 

With a less detailed look, the overall picture of the $\msbar$ masses in
\tab{masses} suggests that -- at the present level of errors -- the b-quark
mass is correctly determined from the different approaches.  Our method is very
different from those which enter the PDG average. It avoids perturbative errors
in all stages of the computation except for the connection of the RGI mass to
the running mass in the $\msbar$ scheme, where truncation errors seem to be
very small.  Due to these properties, it remains of interest to apply our
method with at least three light dynamical quarks and test the consistency of
the table once more. As remarked earlier, the error budget of our present
computation is such that in a future computation a significantly more precise
number can be expected.

\begin{table} 
 \small
 \centering
 \begin{tabular}{l|l@{\hskip 0em}llll|ll}
   \toprule
    $\Nf$ & Ref. & $M$ & $\mbar_\msbar(\mbar_\msbar)$ 
    & $\mbar_\msbar(4\,\GeV)$  & $\mbar_\msbar(2\,\GeV)$ & $\Lambda_\msbar$[\MeV]
    \\\midrule
    0 & \cite{DellaMorte:2006cb}     & 6.76(9)  & 4.35(5)  & 4.39(6)  & 4.87(8)  & 238(19)~\cite{Capitani:1998mq}    
    \\
    2 & \parbox[t]{2cm}{this work}   & 6.58(17) & 4.21(11) & 4.25(12) & 4.88(15) & 310(20)~\cite{Fritzsch:2012wq}
    \\
    5 & PDG13\cite{Beringer:1900zz} & 7.50(8)  & 4.18(3)  & 4.22(4)  & 4.91(5)  & 212(8)\quad~\cite{Beringer:1900zz}
    \\\bottomrule
 \end{tabular}
  \caption{\label{masses}%
           Masses of the b-quark in $\GeV$ in theories with different 
           quark flavour numbers $\Nf$ and for different schemes/scales 
           as well as $\Lambda_\msbar$ and the  RGI mass $M$.
           The PDG value of the b-quark mass is dominated by 
           \cite{Chetyrkin:2009fv, McNeile:2010ji}. 
          }
\end{table}

\begin{appendix}

\section{Error propagation and conversion to $\mbar(\mbar)$}\label{sec:a1}

Here we give details on the conversion function $\rho(r)$ that has been used
in~\eqref{eq:MSconv}. It connects the RGI quark mass $M$ to the quark mass
$m_*$ defined by $\mbar(m_*)=m_*$ and usually denoted by $\mbar(\mbar)$. We
closely follow the standard steps which have been outlined in our notation
in~\cite{Sommer:2010ic}. In a given scheme our conventions for the RG
invariants read
\begin{align}   \label{eq:Lambda}
        \dfrac{\Lambda}{\mu} &=  
        [b_{0}\gbsq(\mu)]^{-\frac{b_1}{2b_0^2}}\,{\rm e}^{-\frac{1}{2 b_{0}\gbsq(\mu)}}
                 \exp\bigg\{
                 \!-\! \int_{0}^{\gbar(\mu)}\!\! {\rm d}g \left[ \frac{1}{\beta(g)}+\frac{1}{b_0 g^3}-\frac{b_1}{b_0^2 g}  \right]
                          \bigg\}\equiv \varphi_g(\gbar)  \;,  \nonumber\\[-1ex]\\
        \dfrac{M}{\mbar(\mu)} &=  
        [2b_{0}\gbsq(\mu)]^{-\frac{d_0}{2b_0}}
                 \exp\bigg\{
                 \!-\! \int_{0}^{\gbar(\mu)}\!\! {\rm d}g \left[ \frac{\tau(g)}{\beta(g)}-\frac{d_0}{b_0 g}  \right]
                          \bigg\} \equiv \varphi_m(\gbar)\,,     \label{eq:Mq}
\end{align}
with universal coefficients $b_0=(11 - 2\Nf/3)(4\pi)^{-2} $, $b_1=(102 -
38\Nf/3)(4\pi)^{-4} $ and $d_0=8 (4\pi)^{-2}$, c.f.~\cite{DellaMorte:2005kg}.
From their ratio one obtains a relation
\begin{align}      \label{}
        r &\equiv \dfrac{M}{\Lambda} = \dfrac{\mbar(\mu)}{\mu}\times 
        \dfrac{\varphi_m(\gbar(\mu))}{\varphi_g(\gbar(\mu))}
\end{align}
that for fixed ${\mbar(\mu)}/{\mu}$ allows us to parameterize the renormalized
coupling $\gbsq(\mu)$ through $r$. Choosing $\mu=m_*$ with $\gbar(m_*)=g_*$ in
eq.~\eqref{eq:Mq} then leads to the functional dependence
\begin{align}
        m_* &= M \cdot\rho(r)\; ,\quad\text{with} \quad
        \rho(r) = 1/\varphi_m(g_*)\,.
        \label{eq:def-rho}
\end{align}
We evaluate this function at 4-loop order in the $\MSbar$ scheme for $\Nf=2$
flavours and obtain to a very good approximation $\rho(r)=0.6400-0.0043\cdot(r-21)$ 
close to $r=21$.

Let us now turn to the propagation of errors from the non-perturbative quark
mass renormalization and coupling renormalization to $\mbar(\mbar)$.  To
incorporate correlations among our non-perturbative data for $M$ and $\Lambda$,
we write
\begin{align}  \label{eq:rho}
  r   &=  \frac{L_0 M}{L_0 \lMSbar} =  \frac{L_1\mbar_\mathrm{SF}(L_0)}{2} \frac{h(L_0)}{k(L_0)} \,,& 
  L_0 &= L_1/2 \,,
\end{align}
where we made use of the definitions 
$z=L_1 M = L_1\,h(L_0)\,\mbar_\mathrm{SF}(L_0) $~\cite{Blossier:2012qu} and
\begin{align}  \label{eq:hk}       
       h(L_0) &= \frac{M}{\mbar_\mathrm{SF}(L_0)}  \,,  &
       k(L_0) &= \Lambda_\mathrm{SF} L_0 \cdot \left[ \dfrac{\Lambda_{\MSbar}}{\Lambda_{\mathrm{SF}}} \right] \,,
\end{align}
with $\Lambda_\msbar / \Lambda_\mathrm{SF} = 2.382035(3)$.  The factors
$h(L_0)$ and $k(L_0)$ are determined by the running of the quark mass and the
coupling in the Schr\"odinger Functional (SF) scheme.  They are known
non-perturbatively in terms of the step scaling functions of
\cite{DellaMorte:2005kg}.  For the error analysis we take the errors in $h,k$
including their correlation into account, remembering that $h$ also contributes
through $\Mb=h(L_0)\mbar^\mathrm{SF}_\mathrm{b}(L_0)$.  The uncertainty arising
from the perturbative running in the $\msbar$ scheme is negligible.  For
example, adding the recently computed 5-loop term in the mass anomalous
dimension \cite{Kostiatalk} does not change numbers at the one permille level.
The error analysis for $\mbar_\msbar(\mu)$ with some fixed $\mu$ is carried
through analogously.

\end{appendix}

\begin{acknowledgement}%
This work is supported in part by the SFB/TR~9,
by grant HE~4517/2-1 (P.F. and J.H.) and HE~4517/3-1 (J.H.),
of the Deutsche Forschungsgemeinschaft,
and by the European Community through
EU Contract MRTN-CT-2006-035482, ``FLAVIAnet''.
It was also partially supported by the Spanish Minister of Education 
and Science projects RyC-2011-08557 (M.~D.~M.). 
P.F. and R.S. thank the ECT* in Trento for support during the workshop
``Beautiful Mesons and Baryons on the Lattice''.
We thank our colleagues in the CLS effort for the joint production
and use of gauge configurations.
We gratefully acknowledge the computer resources
granted by the John von Neumann Institute for Computing (NIC)
and provided on the supercomputer JUROPA at J\"ulich
Supercomputing Centre (JSC) and by the Gauss Centre for
Supercomputing (GCS) through the NIC on the GCS share
of the supercomputer JUQUEEN at JSC,
with funding by the German Federal Ministry of Education and Research
(BMBF) and the German State Ministries for Research
of Baden-W\"urttemberg (MWK), Bayern (StMWFK) and
Nordrhein-Westfalen (MIWF), as well as
within the Distributed European Computing Initiative by the 
PRACE-2IP, with funding from the European Community's Seventh 
Framework Programme (FP7/2007-2013) under grant agreement RI-283493,
by the Grand \'Equipement National de Calcul Intensif at CINES in 
Montpellier under the allocation 2012-056808,
by the HLRN in Berlin, and by NIC at DESY, Zeuthen.
\end{acknowledgement}

\end{document}